\documentclass[prl,aps,twocolumn,groupedaddress,floats,showpacs]{revtex4}
\usepackage{graphicx}
\usepackage{dcolumn}
\usepackage{bm}
\usepackage{color}
\def\he4{$^4$He}
\def\hel3{$^3$He}
\def\Am3{\AA$^{-3}$}
\def\beq{\begin{equation}}
\def\eeq{\end{equation}}

\begin{document}

\author{A. B. Kuklov}
\affiliation{Department of Engineering Science and Physics,
CUNY, Staten Island, NY 10314}

\author{B.V. Svistunov}
\affiliation{Department of Physics, University of Massachusetts,
Amherst, MA 01003, USA}
\affiliation{Russian Research Center ``Kurchatov Institute,''
123182 Moscow, Russia}

\title{Counterflow Superfluidity of Two-Species Ultracold Atoms\\
in a Commensurate Optical Lattice}

\begin{abstract}
In the Mott-insulator regime, two species of ultracold atoms in
optical lattice can exhibit the low energy counterflow motion. We
construct effective Hamiltonians for the three classes of the
two-species insulators---fermion-fermion, boson-boson, and
boson-fermion type---and reveal the conditions when the resulting
groundstate supports super-counter-fluidity (SCF), with the
alternative being phase-segregation. We emphasize a crucial role
of breaking the isotopic symmetry between the species for
realizing the SCF phase.
\end{abstract}

\pacs{PACS 03.75.Fi, 05.30.Jp}

\maketitle

\narrowtext

Recent pioneering experiment by Greiner {\it et al}.
\cite{Greiner} ( proposed in ref.~\cite{Jaksh})
on the superfluid--Mott-insulator transition
in the optical lattice of ultracold atoms has opened
up a new era of strongly correlated atomic systems. Now it is
possible to experimentally study a Mott phase of ultracold atoms.
As shown in
ref.\cite{DEM}, the spinor (S=1) bosons in the lattice can exhibit
very rich and exotic phase diagram.

In the Mott phase of {\it single} specie of atoms, any
low-energy transport is suppressed \cite{MIT}. The situation
may change dramatically, if the Mott phase includes al least {\it
two} species. While the net number-of-atoms transport
is still suppressed, the counterflow (when the currents of
the two species are equal in absolute values and are in opposite
directions), generally speaking, survives, and, at certain
conditions, it can be non-dissipative (super-counterflow).

In this Letter, we discuss typical cases of counterflow dynamics
in two-species Mott phase. We consider three classes of the
commensurate two-species systems: fermion-fermion (FF),
boson-boson (BB), and boson-fermion (BF) mixtures (assuming that
each of the species does not possess internal degrees of freedom).
We confine our analysis to the strong coupling limit, when the
effective counterflow Hamiltonian is readily obtained in the
second order perturbation theory in the parameter $t/U\ll 1$, where $t$
is the hopping amplitude and $U$ is the on-site repulsion. For the
same-statistics (BB or FF) mixtures, the effective Hamiltonian can
be written in terms of $s=\nu/2$ iso-spin operators (the integer
$\nu$ is the filling factor). In the FF case, we have a
$s=1/2$ anti-ferromagnetic asymmetric effective Hamiltonian.
In the BB case, the counterflow term
is represented by asymmetric ferromagnetic interaction of iso-spins.

Depending on particular means of breaking the isotopic ~$SU(2)$
symmetry, either the effective
Hamiltonians are of the easy-axis type, which means
phase segregation, or they are of the easy-plane type resulting in
the SCF groundstate. In the BF case,
it is possible to eliminate bosonic
variables and obtain a spinless fermionic model with
nearest-neighbor interaction. The sign of the interaction
depends on the parameters of the original Hamiltonian, and how
they deviate from the exact fermion-boson symmetry. In particular,
the $p$-wave Cooper pairing is possible which results
in the SCF.

Our lowest-order effective Hamiltonians will involve only on-site
and nearest-neighbor interactions. Hence, to derive them it is
sufficient to consider a system consisting of two strong traps, 1
and 2, between which weak tunneling is allowed. Only the
lowest one-particle state ~$\varphi_{\sigma}$~ for the species
~$\sigma =\uparrow, \downarrow$ (we use pseudo-spin notation) in
each trap is taken into account. We introduce
the creation and annihilation fermionic/bosonic operators
~$a^{\dagger}_{b\sigma},\,\, a_{b\sigma}$, respectively, where
~$b=1,2$ labels the trap (site). Our original
two-site Hamiltonian then reads
\begin{eqnarray}
\displaystyle H_{12}&=&H^{(0)}_{12} + V_{12} \; ,
\label{HAM}
\\
H^{(0)}_{12}&=&{1 \over 2} \sum_{b \, \sigma \sigma'} \, U_{\sigma
\sigma'} \, \mbox{:} \, n_{b\sigma} n_{b\sigma'} \mbox{:} \; ,
\label{HO}
\\
V_{12}&=& -\sum_{\sigma} \, (t_{\sigma} a^{\dagger}_{1\sigma}
a_{2\sigma} + H.c.) \; ,
\label{V12}
\end{eqnarray}
\noindent where ~$n_{b\sigma}= a^{\dagger}_{b\sigma} a_{b\sigma}$,
$U_{\sigma \sigma'} = g_{\sigma \sigma'}\int d{\bf
x}|\varphi_{\sigma}|^2 |\varphi_{\sigma'}|^2$, $g_{\sigma
\sigma'}$ is the interaction constant between species $\sigma$ and
$\sigma'$, and $\mbox{:} (\ldots) \mbox{:}$ denotes normal form of
the product of the creation-annihilation operators.
We consider the
tunneling matrix element ~$t_{\sigma}$~ as real and positive
and depending
on the component index.

At commensurate total filling ~$n_b = \sum_{\sigma}
n_{b\sigma}=\nu$~ on each site, and in the limit ~$|t_{\sigma}|\ll
U_{\sigma \sigma'}$, single particle jumps change the total
on-site populations, and, therefore, require high energy. In
contrast, exchanging two different particles does not require such
an energy. In order to describe these processes the single
particle jumps should be eliminated in the second order with
respect to the ratio ~$t/U$. Thus, we choose ~$H_{12}^{(0)}$~ in
eq.~(\ref{HO}) as zeroth order part, while ~$V_{12}$~ is the
perturbation. The basis in our effective Hilbert space is given by
the states ~$|{\rm eff} \rangle= |n_{1\uparrow}, n_{1\downarrow};
n_{2\uparrow}, n_{2\downarrow}\rangle$, where ~$ n_1= n_2=\nu \,$;
the virtual excited states ~$|{\rm ex} \rangle $~ (where ~$n_1=n_2
\pm 1$) are excluded in the second order. The effective
Hamiltonian ~$V_{12}'$ can be represented in terms of the matrix
elements of the original Hamiltonian as
\begin{eqnarray}
\displaystyle (V_{12}')_{\alpha \beta} =-\sum_{\gamma} {V_{\alpha
\gamma} V_{\gamma \beta} \over E_{\gamma} - (E_{\alpha} +
E_{\beta})/2 } \; ,
\label{V'}
\end{eqnarray}
\noindent where ~$\alpha,\,\,\beta$~ label ~$|{\rm eff} \rangle$,
and ~$\gamma$~ denotes ~$|{\rm ex} \rangle$; ~$E$'s~ are the
eigenenergies of ~$H^{(0)}$~ in eq.~(\ref{HO}).

{\large \it FF case}. This case is the easiest for calculation.
The only possible filling factor is $\nu=1$ and the only possible
intermediate virtual state $\gamma$ is the state with two
different fermions on one site. The only relevant interaction
vertex is $U_{\uparrow \downarrow} = U$. The Hamiltonian of the
system is the standard Hubbard model
\cite{Hubbard} widely used for electrons in metal, the only
specific (and significant !) feature being the dependence of the
hopping amplitude on the ``spin" index. The effective
Hamiltonian is readily obtained from eq.~(\ref{V'}), in terms of
iso-spin operators
\begin{equation}
{\bf S}_b = (1/2)\sum_{\sigma \sigma'} \, a^{\dag}_{b \sigma}
({\bf \sigma})_{\sigma \sigma'} a_{b \sigma}\; ,
\label{iso}
\end{equation}
where $({\bf \sigma})_{\sigma \sigma'} $~  are the Pauli
matrices. The operators  ${\bf S}_b=(S_{bx}, \, S_{by},\, S_{bz})$,
where the second index
refers to the iso-spin components, obey
standard spin commutation relation at same site
and commute with each other on different sites.
The resulting effective two-site
Hamiltonian ~$ H_{12}'= 2J{\bf S}_1{\bf S}_2 + 2J'S_{1z}S_{2z}$~
extended
to a lattice is

\begin{eqnarray}
\displaystyle H_{FF}&=& \sum_{\langle ij \rangle} \, [  J{\bf
S}_i{\bf S}_j + J'S_{iz}S_{jz}] -\sum_i \, B S_{iz} \; ,
\label{HFF} \\
J&=&2t_{\uparrow}t_{\downarrow} / U \; , ~~~~~ J'=(t_{\uparrow} -
t_{\downarrow})^2 / U \; .
\label{J}
\end{eqnarray}
\noindent where $\langle \ldots \rangle $ stands for the pairs of
nearest neighbors; ~$B=\mu_{\uparrow}-\mu_{\downarrow}$, and
~$\mu_\sigma$~ is the chemical potential of the $\sigma$-th
component. If the two species are,
e.g., just different isotopes of one and the same element, then
the condition $t_{\uparrow} \neq t_{\downarrow}$
naturally occurs from the difference of the masses.

The groundstate of this system  (see, e.q., \cite{textbook}) at
~$B=0$~ is the spin-1/2 easy-axis anti-ferromagnet, given
~$J>0,\,\, J'>0$ in eq.~(\ref{J}). This corresponds to the
checkerboard insulator (N\'eel state). If ~$J'=0$~ and for
infinitesimally small ~$B$, this state is unstable with respect to
forming the {\it canted} phase, that is, the easy-plane state in
which the staggered magnetization resides in the $(x,y)$-plane,
with the ferromagnetic component ~$\sim B$~ oriented along the
~$z$-axis. This state maps into the superfluid bosonic system (see
in, e.g., \cite{SACHDEV} and also \cite{0110024}),
which is equivalent to the SCF.
For
~$J'\neq 0$, the {\it canted} state occurs by the first order
transition at ~$|B|>B_c$, where ~$B_c\approx \sqrt{J'(2J+J')}$.
In
the {\it canonical} ensemble, when the numbers of atoms of both
components, $N_{\uparrow}$ and $N_{\downarrow}$, are fixed, the
SCF phase corresponds to a {\it general} case, while the
checkerboard insulator occurs only at $N_{\uparrow} =
N_{\downarrow}$. As shown in ref.\cite{0110024}, in canonical
ensemble, the checkerboard insulator state can coexist with
the superfluid, with the amount of the superfluid determined
by ~$N_{\uparrow} -N_{\downarrow}\neq 0$.

It is important to stress, that in our situation---in contrast to
the case of real easy-plane anti-ferromagnet---the staggered
magnetization vector fundamentally cannot be attached to any
direction in the easy plane due to the exact conservation of the
atoms of each component. This guarantees the exact $U(1)$
symmetry. As discussed later, the easy-plane $U(1)$ symmetry can
be broken only {\it spontaneously} by forming the superfluid
counter-flow vacuum.

{\large \it BB case}. In this case we also introduce the site
iso-spin operators according to eq.~(\ref{iso}) with bosonic
creation-annihilation operators. The expression for the effective
Hamiltonian in a general case of non-equal interactions is rather
cumbersome. It becomes much simpler in two important particular
cases: (i) when ~$U_{\sigma \sigma'} = U + \delta U_{\sigma
\sigma'}$, with ~$|\delta U_{\sigma \sigma'}| \ll U$, and (ii)
when $\nu=1$. We start with the former case in the limit ~$|\delta
U_{\sigma \sigma'}|/ U \to 0$. Then, the denominators in
eq.~(\ref{V'}) are the same and equal to ~$U$. Hence, the
intermediate virtual states $\gamma$ do not involve any projection
operators, and the answer for the two-sites can be written as a
bilinear form of $s=\nu/2$ iso-spin operators as ~$ H_{12}'=-
J{\bf S}_1{\bf S}_2 - J'S_{1z}S_{2z} -B_{12}(S_{1z} + S_{2z}) \; ,
B_{12}=(\nu+1) (t^2_{\uparrow}- t^2_{\downarrow}) /U$, where the
expressions for $J$ and $J'$ are given by eqs.~(\ref{J}). When $
|\delta U_{\sigma \sigma'}| \ll U$, the parameters $J$, $J'$, and
$B_{12}$ remain, to the first order in  $|\delta U_{\sigma
\sigma'}|/U$, the same. The main correction to the Hamiltonian is
associated with the term ~$H^{(0)}_{12}$~ (\ref{HAM}). Then,
extending the two-sites Hamiltonian ~$H'_{12}+ H^{(0)}_{12}$~ to
the lattice and omitting trivial constant terms, we find

\begin{eqnarray}
\displaystyle H_{BB}&=&-\sum_{\langle ij \rangle} \, [ J{\bf
S}_i{\bf S}_j + J'S_{iz}S_{jz}] \nonumber \\ &+&\sum_i \, [ (D/2)
(S_{iz})^2 - B S_{iz}] \; ,
\label{HBB} \\
B&=& \mu_{\uparrow}-\mu_{\downarrow} -\frac{1}{2}(\nu-1)(U_{\uparrow
\uparrow}-U_{\downarrow \downarrow})+ p \, B_{12} \; ,
\label{B} \\
D&=&U_{\uparrow \uparrow}+U_{\downarrow \downarrow}-2U_{\uparrow
\downarrow} \; ,
\label{cond}
\end{eqnarray}
\noindent
with ~$p$~ being the number of close neighbors on the lattice.
Note that in contrast to the FF case,
the minus sign now stands in
front of ~$J$ and $J'$. That is, if the ~$D$-term
in eq.(\ref{HBB}) is ignored, we arrive at the easy-axis
ferromagnetic model, that implies a phase
segregation. However, for large enough ~$D$~ in eqs.(\ref{HBB},\ref{cond}),
the easy-plane ground state can be realized.
The mean field condition for this (in the
case ~$B=0 \,$) is ~$D\gtrsim pJ'$ \cite{NOTE5}.

The effective Hamiltonian for the BB
situation at $\nu=1$ can readily be found for arbitrary $U_{\sigma
\sigma'}$. The summation over $\gamma$ in eq.(\ref{V'})
in this case is very
simple, because for any different states $\alpha$ and $\beta$ in
eq.~(\ref{V'}), there is no more than one state $\gamma$ for which
matrix elements differ from zero. In the diagonal terms $\alpha =
\beta$, if two sites are both occupied by the same
iso-spin ~$\sigma$~ bosons, the energy change due to jumping of
either boson is ~$U_{\sigma \sigma}$; when the bosons have
opposite spins, the energy change becomes ~$U_{\uparrow
\downarrow}$. The final result
acquires the form eq.~(\ref{HBB}), the parameters $J$, $J'$, and $B$
being
\begin{eqnarray}
\displaystyle J&=&2t_{\uparrow}t_{\downarrow} / U_{\uparrow
\downarrow} \; ,
\label{a} \\
J'&=& - (t_{\uparrow} +
t_{\downarrow})^2 / U_{\uparrow \downarrow} +2t^2_{\uparrow} /
U_{\uparrow \uparrow} +2t_{\downarrow}^2/ U_{\downarrow
\downarrow} \; ,
\label{b} \\
B_{12}&=&
2 (t^2_{\uparrow} / U_{\uparrow \uparrow }-
t^2_{\downarrow} / U_{\downarrow \downarrow} ) \; .
\label{c}
\end{eqnarray}
At certain conditions, one obtains the easy-plane situation $J'<0$,
that is the SCF. A solution for this can, in principle, be found
exactly. We will, however, analyze it in a simplified situation
when the system is almost $SU(2)$ symmetric (like $^{87}$Rb
\cite{RB}), which means ~$U_{\uparrow \uparrow }\approx
U_{\downarrow \downarrow} \approx U_{\uparrow\downarrow},\,\,
t_{\downarrow}\approx t_{\uparrow}$, and will choose chemical
potentials to have $B=0$. We represent ~$U_{\uparrow
\uparrow}=U+U'+D/2,\,\, U_{\downarrow \downarrow}=U-U'+D/2,\,\,
U_{\uparrow\downarrow}=U$, ~$t_{\uparrow}=t+t',\,\,
t_{\downarrow}= t-t'$, with ~$|U'|/U\ll
1,\,\, |D|/U\ll 1,\,\, |t'/t|\ll 1$. Then, ~$J'<0$~ yields

\begin{eqnarray}
\displaystyle
|t'/t|< \sqrt{D/2U},\,\, D>0.
\label{AF}
\end{eqnarray}

{\large \it The easy-plane state is equivalent
to the SCF}.
Now let us focus on why the easy-plane ground state
of the models (\ref{HFF}, \ref{HBB})
can support the superfluid counter-flow of the
components.
First, we note that, according
to Holstein and Primakoff \cite{HP},
spin lattice is equivalent to the lattice
bosons, with the spin commutation relation being
essentially equivalent to the Bose commutation
relation. Then, the site operators ~$\hat{S}^+_n=\hat{S}_{nx}
+ i\hat{S}_{ny}$~
are proportional to the effective boson annihilation
operators ~$\hat{b}_n$~ \cite{HP}, and the operator
~$\hat{S}_{nz}=(\nu /2) - \hat{b}^\dagger_n\hat{b}_n$.
Accordingly, the formation
of the easy-plane order parameter
~$ S^+_n =\langle S_{nx} + iS_{ny}\rangle=
|S'|\exp (i\varphi)\sim \langle \hat{b}_n\rangle$, where
~$|S'|\approx const \neq 0$, is equivalent
to the formation of the Bose-Einstein condensate (BEC)
~$\langle \hat{b}_n\rangle \neq 0$~
of the Holstein-Primakoff boson ~$\hat{b}_n$.
As long as the global invariance with respect to the
phase ~$\varphi$~ of the bosons holds, which
is insured by the conservation
of the original species, and the bosons are interacting,
the system is (counter-) superfluid.

Let us obtain expression for the superfluid counter-current
density ~$\bf I$, which is the difference of the currents of the
each specie. To be specific, we will discuss the BB case and keep
in mind that the result for the FF case is, qualitatively, the
same. In the long wave limit, the conservation law of the
components takes the form ~$\dot{S}_z + {\bf \nabla}{\bf I}=0$,
where ~$ S_z({\bf x})\approx S_{nz}/\Omega $~
is the difference of the densities of
the species, with ~$\Omega$~ standing for the unit cell volume
and ~$\bf x$~ being close to the site $n$.
This continuity equation follows from the Heisenberg equation of
motion ~$i\dot{S}_{nz} =[S_{nz}, H_{BB}]$~ after performing the
commutations and taking the long wave limit. Finally, replacing
the operators by $c$-numbers, we find

\begin{eqnarray}
\displaystyle {\bf I}= n_{\rm scf}{\bf \nabla} \varphi ,
\,\,\,\,\, n_{\rm scf}= Jp \, d^2|S'|^2 /(12\Omega),
\label{I}
\end{eqnarray}
\noindent where ~$d$~ denotes the nearest neighbor distance;
~$n_{\rm scf}$~ has a meaning of the effective superfluid density,
with ~${\bf \nabla} \varphi$~ being the corresponding velocity.
Thus, the easy-plane ground state of the models (\ref{HFF},
\ref{HBB}) supports a super-counterflow of the two components. The
general condition for this is that the intrinsic symmetry (i.e.,
$SU(2)$) between the components is broken down to the $U(1)$
group.

{\large \it BF case}. Now we turn to the Bose-Fermi mixture on the
lattice. The Hamiltonian (\ref{HAM}) of two sites becomes

\begin{eqnarray}
\displaystyle H^{(0)}_{12}&=&\sum_{b=1,2}[ (U_0 / 2) \, n_b(
n_b-1) +U_1 \, n_bm_b]\; ,
\label{HOF}
\\
V_{12}&=&-(t_B \, a^{\dagger}_2a_1  + t_F \, c^{\dagger}_2c_1
+H.c.) \;  ,
\label{V12F}
\end{eqnarray}
\noindent
where ~$a^{\dagger}_b,\,\,a_b$~ and
~$c^{\dagger}_b,\,\,c_b$~ stand for the site creation and annihilation
operators of the bosons and the fermions,
respectively; $n_b= a^{\dagger}_b a_b, \; m_b= c^{\dagger}_b c_b$;
~$U_0>0,\,\, U_1>0$.
Below we will see that the SCF is possible, if (formal)
symmetry between bosons and fermions (when ~$U_0=U_1,\,\, t_B=t_F$)
is broken.

The effective two-site Hamiltonian can be written in terms of the
fermionic operators only. The elimination of the bosonic operators
is possible because the truncated Hilbert space is exhausted by
different fermionic occupations, the bosonic occupations being
unambiguously defined by the constraint $n_b+m_b=\nu$. Formally,
one introduces new operators ~$\tilde{c}_i=a^\dagger_ic_i/
\sqrt{\nu}$~ and observes that, given our constraint, these
satisfy standard fermionic algebra. [Note, that the vacuum for the
new fermions corresponds to the Bose Mott insulator state with the
filling factor $\nu$.] The effective Hilbert space and the
Hamiltonian are then automatically defined:


\begin{eqnarray}
\displaystyle H_{BF}=\sum_{\langle ij \rangle} \, [-\kappa
c^{\dagger}_ic_j + (\lambda / 2) \, m_i m_j]- \mu_F \sum_i \, m_i
\; ,
\label{HBF}
\end{eqnarray}
\noindent (we have omitted the tilde in the fermionic operators).
Here ~$\mu_F= \mu + p \, \mu_{12}$~ denotes the resulting
fermionic chemical potential, with ~$\mu$ being the difference of
the bare fermion and boson chemical potentials (adding one fermion
to the lattice implies simultaneous removal of one boson); the
other constants are:

\begin{eqnarray}
\kappa &=& 2 \, \nu \,  t_Bt_F / U_1 \; ,
\label{KAP}\\
\lambda &=& 2t_B^2 \left({\nu^2\over U_1} + {\nu^2 - 1\over 2U_0-U_1}-
{2\nu^2\over U_0} \right) +{2t_F^2 \over U_1} \; ,
\label{LAM}\\
\mu_{12} &=&  {t_F^2+t_B^2 \nu^2 \over U_1} + {t_B^2(\nu^2-1)\over
2U_0-U_1} - {2t_B^2 \nu(\nu+1) \over U_0} \; ,
\label{MU}
\end{eqnarray}
and it is assumed that~$2U_0 - U_1>0$~ \cite{NOTE2}.

Note that in the case of the exact boson-fermion symmetry (in the
meaning defined above), ~$\lambda =0$~ and SCF is not possible,
because the problem maps on the ideal one-component fermion gas.
Thus, the symmetry must be broken. From eqs.~(\ref{LAM}), it is
seen that there is a region where $\lambda <0$. At, e.g.,
~$U_1=U_0$, we find ~$\lambda= 2(t_F^2 -t_B^2)/U_0$,
and ~$\lambda <0$~ when ~$t_B>t_F$. Thus, we obtain
one-component lattice fermions with the nearest neighbor
attraction. This leads to
the Cooper pairing (see, e.g., \cite{LP}),
which in our case implies the SCF. As our fermions are effectively
spinless (spin-polarized), the pairing takes place in the
$p$-channel, that is the superconducting order parameter is
characterized by broken inversion symmetry, and can exhibit broken
time-reversal symmetry (see in \cite{SIG}). The particular structure
depends on the lattice.

{\large \it Experimental realization and detection of the SCF}.
The essential condition for the SCF is that the system is in the
MI regime with respect to the net-atomic transport. It is exactly
the same condition formulated for the one component case
\cite{Jaksh}. In the FF case at
~$T=0$~ and ~$|B|> B_c$, the system will be in the SCF state.
In the BB
case, the three interaction constants ~$U_{\sigma \sigma}$~ must
be tuned in order to avoid the phase separation. This, however,
can be achieved by creating slightly displaced lattice potentials
\cite{Jaksh} for the species in order to reduce the overlap of the
states ~$\uparrow ,\, \downarrow$~ on each site, so that the value
~$U_{\uparrow \downarrow}$~ is reduced and the parameter ~$D$~
(\ref{cond}) becomes positive and large enough to insure the
condition (\ref{AF}) for, e.g., $^{87}$Rb.

It is important to note that in the cases FF and BB, once ~$T\neq
0$~ the phase transition from the ordered (SCF) phase into the
``paramagnetic" (i.e., normal with respect to the counter-flow)
state may occur. A typical critical (N\'eel or Curie) temperature
~$T_c$~ is given by the effective exchange constant as
~$T_c\approx pJ$. Thus, the SCF requires low enough temperatures
~$T\leq pt^2/U$. For typical parameters employed in
ref.\cite{Jaksh}, this reads as ~$T\leq 10^{-9}$K.  In the BF
case, the SCF-normal counter-fluid transition temperature is given
by the same energy scale.

The SCF could be detected by observing
non-dissipative exchange of the components
through the optical lattice. One of the possibilities
consists of separating the components initially.
Then, the species will
be exchanging their position in the oscillating manner.
Raising ~$T$~ above
~$T_c$~ will result in the abrupt increase of the
damping of these oscillations.

The existence of the phase ~$\varphi$~ (\ref{I})
implies that the counter-vortex can be
supported in the SCF phase. Then, the winding
of the phase in the SCF, which has a boundary
with the regular BEC phase of the BB components,
will imprint this winding on the BEC phases, resulting
in regular BEC vortices. Detecting these vortices will, then,
signal a presence of the SCF-vortex.

The order associated with the SCF is revealed in the {\it
two}-particle density matrix. Its detection can be done by
atomic scattering of fast atoms off the lattice \cite{ATOM}.
Specifically, the cross-section
of the process, when incoming fast particle of, e.g., sort
~$\uparrow$~ strikes the atoms in the lattice and ``transforms"
into the outcoming particle ~$\downarrow$, is exactly given by the
(iso) spin-spin correlators
of the atoms in the lattice. This issue will be considered in
greater detail elsewhere.

Finally, let us discuss a formation of the SCF vortex in a
rotating lattice. To be specific, we will discuss the BB case. The
lattice rotation can induce such vortices, if the components have
different masses ~$M_\sigma$. Indeed, the total mass current can
be expressed as ~${\bf I}^{(M)}= (M_\uparrow-M_\downarrow)n_{\rm
scf}{\bf \nabla}\varphi $, where ~$n_{\rm scf},\,\, \varphi$
define the counter-current (\ref{I}). External rotation of the
lattice at some angular velocity ~$\omega$ changes the energy of
the counter-vortex ~$\sim n_{\rm scf} L\ln (L/d)$, where $L$
stands for a typical system size, by ~$\sim \pm
(M_\uparrow-M_\downarrow ) n_{\rm scf} L^3\omega$.
The formation of the vortex becomes energetically favorable at
\cite{LP}

\begin{eqnarray}
\displaystyle
|\omega|\geq \omega_c \approx L^{-2}\ln (L/d)/|M_\uparrow-M_\downarrow|.
\label{CR}
\end{eqnarray}
\noindent

Note that differences in masses
can result in the phase separation induced by gravity/rotation.
However, if the atomic magnetons of different species
are different, the corresponding linear and oscillator
fields can be compensated by adding non-uniform
magnetic field.

Summarizing, we have demonstrated that there is a large variety of
strongly correlated groundstates in the two-component system of
ultracold atoms in an optical lattice, even if the system is in
the Mott-insulator regime with respect to the net number-of-atoms
transport. The most dramatic effect that may occur
is the so-called super-counter-fluidity, when the
system supports a non-dissipative counterflow of the two
components. In the strong-coupling limit, the effective
Hamiltonian for the super-counterfluids corresponds to: (i)
easy-plane ({\it canted}) anti-ferromagnet (in the case of fermion-fermion
mixture), (ii) easy-plane ferromagnet (in the case of boson-boson
mixture), and (iii) $p$-wave superconductor (in the case of
boson-fermion mixture). The experimental conditions
and means of detecting the super-counter-fluidity are
outlined.

Authors acknowledge useful discussions of the results with David
Schmeltzer, Nikolay Prokof'ev and Manfred Sigrist.
ABK was supported by CUNY grant
PSC-63499-0032. BVS acknowledges a support from Russian Foundation
for Basic Research under Grant 01-02-16508, from the Netherlands
Organization for Scientific Research (NWO), and from the European
Community under Grant INTAS-2001-2344.


\begin{references}
\bibitem{NOTE2}
Otherwise, for ~$\nu >1$, the ground state of (\ref{HOF})
will have $\nu -1$ ~ particles on a site, containing
fermion, and
~$\nu +1$ bosons on the other site with no fermion. In this
situation, no transport of the  fermion is possible in the second
order in ~$t_B,\,\, t_F$.
\end{references}
\end{document}